\documentclass[preprint]{aastex}

\slugcomment{Accepted for publication in The Astrophysical Journal}

\shorttitle{Chemical equilibrium and stable stratification}
\shortauthors{Reisenegger}

\begin{document}

\title{Chemical equilibrium and stable stratification of a multi-component fluid:
thermodynamics and application to neutron stars}

\author{Andreas Reisenegger}
\affil{Departamento de Astronom\'\i a y Astrof\'\i sica, 
Pontificia Universidad Cat\'olica de Chile, \\
Casilla 306, Santiago 22, Chile\altaffilmark{1}\\
and\\
Institute for Theoretical Physics, University of California,
Santa Barbara, CA 93106-4030, USA}

\email{areisene@astro.puc.cl}

\altaffiltext{1}{Permanent address.}

\begin{abstract}
%A fluid in a gravitational field is said to be {\it stably stratified} (or
%{\it convectively stable}) if
%a fluid element, displaced from its equilibrium position in a direction
%parallel to gravity, is forced by buoyancy forces to return to its initial
%position. This imposes the condition that the mass density 
%(or, more generally, the relativistic enthalpy density) 
%in the perturbation be a weaker function of pressure than the ambient density 
%in the equilibrium fluid.
%If the perturbation occurs fast enough to conserve entropy 
%and chemical abundances, the stability condition can be written in terms 
%of gradients of these quantities in the equilibrium 
%fluid (Ledoux criterion). 
A general thermodynamic argument 
shows that multi-component matter in full chemical equilibrium, 
with uniform entropy per baryon, is generally stably stratified. 
This is particularly relevant for neutron stars, in which the effects of
entropy are negligible compared to those of the equilibrium composition 
gradient established by weak interactions. It can therefore be asserted
that, regardless of the uncertainties in the equation of state of dense
matter, neutron stars are stably stratified. This has important, previously 
discussed consequences for their oscillation modes,
magnetic field evolution, and internal angular momentum transport.
\end{abstract}


\section{Introduction}

Whether a fluid is stably stratified or unstable to convective overturn 
plays a crucial role in its dynamics and transport properties. 
This is particularly true in non-degenerate stars, where heat may be 
transported by convection or by radiation,
and which of the two mechanisms is effective determines the shape of the
temperature profile, and with it much of the structure of the star (see,
e.g., \cite{Kippenhahn}). 
%It also affects the possibility and detailed 
%properties of differential rotation in the star and its effect on a 
%potential stellar dynamo (e.g., \cite{Spruit}).

In the case of neutron stars, this issue had been largely ignored until it was
pointed out that the transport of magnetic flux could be strongly affected
if the matter was stably stratified (\cite{Pethick,GR92}).
Simple models for the composition and equation of state of equilibrium
neutron star matter generally show that it is stably stratified
due to the presence of a chemical gradient, and give more or less 
consistent estimates for the characteristic (Brunt-V\"ais\"al\"a) 
frequency (\cite{Pethick,RG92,Lai,Lee}). 

Stable stratification leads to a set of g-modes (\cite{RG92,Lai}) which might 
be excited during binary inspiral (\cite{RG94,Lai,Ho}). 
It also may affect angular momentum transport by inhibiting Ekman circulation 
(\cite{R93,R95,Abney1}),
with possible observational signatures in pulsar glitches (\cite{Abney2}).
Its effects on magnetic flux evolution have also gained relevance
with the interpretation of soft gamma-ray repeaters
and anomalous X-ray pulsars as ultramagnetized neutron stars powered by 
magnetic field decay (\cite{Thompson1,Thompson2,Kouveliotou,Gotthelf}). 
The recent work on r-modes destabilized by 
gravitational wave emission in very rapidly rotating neutron stars 
(\cite{Andersson,Lindblom}) is also paying increasing attention
to the effects of stable stratification and the relation between the modes 
of stratified and unstratified stars (e.g., \cite{Lockitch}).

The previous work has made it plausible that cold neutron star matter is 
always stably stratified, but it has only considered specific models for
neutron star matter, which are still quite uncertain (\cite{EOS}). 
The present paper shows by a simple thermodynamic argument 
%(first given by the author in a seminar at the
%Institute for Advanced Study in 1994) 
that an isentropic, multi-component 
fluid in full chemical equilibrium cannot be unstable to 
convection.\footnote{A different argument was recently given by 
\cite{Thompson3}.} Since cold neutron stars are zero-entropy objects, this 
proves the point above, except for the remaining, but unlikely possibility
of marginal stability (discussed below). 

In the next section, the stage is set by a heuristic derivation of the
general condition for stable stratification, given previously by
other authors, but now phrased in the language and notation to be
used in the rest of this work. Section 3 contains the thermodynamic proof 
that is the core of the present paper. 
Implications of this result are discussed in the final section. 

\section{Stable stratification}

The stability condition for a static fluid in a gravitational 
field\footnote{In the case of a rotating star with a simple rotation
law, the gravitational field and the corresponding potential can be
generalized to include centrifugal effects.}
$\vec g=-\nabla\phi$
can be obtained by the following heuristic argument (e.g., \cite{Cox}). 
The (Newtonian) condition of hydrostatic equilibrium, 
\begin{equation}
\nabla P=-\rho\nabla\phi,
\end{equation}
requires the gravitational potential $\phi$, the pressure $P$, and the mass 
density $\rho$ in the equilibrium configuration of the fluid to be constant
on the same set of surfaces. Take a fluid element and displace it 
perpendicularly to these surfaces, on a timescale slow 
enough to keep it in pressure equilibrium 
with the surrounding matter, but not necessarily slow enough to 
exchange heat and particles with its surroundings or reach chemical
equilibrium by internal processes. Therefore, the fluid element will
generally reach a different density than the surrounding matter, and 
a buoyancy force will either increase or decrease its initial displacement.
The buoyancy force per unit mass per unit displacement (taken to be positive 
if it acts so as to restore the fluid element to its initial position)
is the squared Brunt-V\"ais\"al\"a frequency,
\begin{equation}
N^2=\left[\left(d\rho\over d P\right)_{eq}
-\left(d\rho\over d P\right)_{pert}\right]|\vec g|^2,
\end{equation}
where the subscripts ``$eq$'' and ``$pert$'' refer, respectively, to variations
in the background equilibrium fluid and in the perturbed fluid element. 
If $N^2>0,$
the fluid can support oscillations with buoyant restoring forces (gravity waves)
and characteristic frequency $N$. If $N^2<0$, the fluid is unstable to 
convective overturn. 

Take the entropy per baryon of the equilibrium fluid to be $S(P)$, 
and its chemical composition to be parameterized by a set of 
variables $Y_i(P)$ (abundance, per baryon, of particle species $i=1,2,..., n$).
The pressure is used to mark the depth in the fluid at which the variables 
are to be evaluated. Assume that the displaced fluid element does not
change its internal variables $S$ and $\{Y_i\}_{i=1}^n$. 
The condition for stable stratification, $N^2>0$, can
then be rewritten in the alternative forms
\begin{equation}
\left(d\rho\over d P\right)_{eq}
-\left(\partial\rho\over\partial P\right)_{S,\{Y_i\}}>0
\end{equation}
or (\cite{Ledoux,Epstein})
\begin{equation}
\left(\partial\rho\over\partial S\right)_{P,\{Y_i\}}
\left(d S\over d P\right)_{eq}
+\sum_i\left(\partial\rho\over\partial Y_i\right)_{P,S}
\left(d Y_i\over d P\right)_{eq}>0,
\end{equation}
where, in the usual thermodynamic notation, the variables in the subscripts 
are to be held constant when taking the partial derivatives. If the sum
involving the composition gradient can be neglected (as usually the case 
in non-degenerate stars such as the Sun and other main-sequence or giant 
stars), the condition reduces to the familiar Schwarzschild 
condition (\cite{Schwarzschild}) of decreasing entropy per baryon with increasing 
pressure.

\section{Thermodynamic proof}

In what follows, it is assumed that the fluid is {\it isentropic} (i.e.,
$S=S_0=$ constant throughout) and that in the unperturbed configuration the matter 
is {\it fully catalyzed}, i.e., the parameters $\{Y_i(P,S_0)\}$ are those corresponding
to full chemical (and thermodynamic) equilibrium at the local pressure. 
This equilibrium state is
determined by minimizing the enthalpy per baryon, $h(P,S_0,\{Y_i\})$ over all
values of $\{Y_i\}$ allowed by conservation laws,
\begin{equation}
h_{eq}(P,S_0)=\min_{\{Y_i\}}h(P,S_0,\{Y_i\}).
\end{equation}
At given entropy per baryon $S_0$, the resulting function $h_{eq}(P,S_0)$ 
defines a curve on the pressure-enthalpy plane (see Fig. 1).
%$(P,h)$ space, in which we take $h$ to be represented along the vertical axis.
Point $I$ on the curve represents the initial (equilibrium)
state of the fluid element to be perturbed. The perturbation, at constant 
$S=S_0$ and $\{Y_i\}$, takes the fluid to a different pressure and {\it out of
equilibrium}, therefore to a point {\it above} the equilibrium curve.
The trajectory of the fluid element is
tangent to the equilibrium curve at the point representing the equilibrium 
state of the fluid element, 
\begin{equation}
\left({\partial h\over\partial P}\right)_{S,\{Y_i\}}
=\left({\partial h\over\partial P}\right)_{S,eq},
\end{equation}
and curves above the equilibrium line to both sides of this point,
\begin{equation}
\left({\partial^2 h\over\partial P^2}\right)_{S,\{Y_i\}}
-\left({\partial^2 h\over\partial P^2}\right)_{S,eq}
=\sum_{i,j}\left({\partial^2 h\over\partial Y_i\partial Y_j}\right)_{S,P}
\left({\partial Y_i\over\partial P}\right)_{S,eq}
\left({\partial Y_j\over\partial P}\right)_{S,eq}\geq 0,
\end{equation}
where the identity can be obtained from standard thermodynamic
relations, together with the minimum condition on $h$ in equilibrium.
The inequality becomes an {\it equality} only in the special cases 
when either the equilibrium concentrations $Y_i$ do not depend on pressure, or
the enthalpy is insensitive to changes in these concentrations. 

We note that $v\equiv(\partial h/\partial P)_{S,\{Y_i\}}$
is the {\it volume per baryon}, so equation (7) can be interpreted as 
a condition on the derivatives of $v$ with respect to pressure: 
\begin{equation}
\left({\partial v\over\partial P}\right)_{S,\{Y_i\}}
\geq\left({\partial v\over\partial P}\right)_{S,eq}.
\end{equation}
In the non-relativistic limit, the mass density $\rho=m/v$, where
$m$ is the (constant) baryon mass, so equation (8) is equivalent to the 
condition for convective stability or marginal stability (eq. 3) for an 
isentropic equilibrium state.
In the more general case of a relativistic fluid, the gravitating mass 
is the enthalpy (including rest mass; e.g., \cite{MTW}), rather than the
rest mass alone. Therefore $\rho$ should be interpreted as the enthalpy
per unit volume, $\rho=h/v$. Since, to lowest order, the changes in $h$ 
are the same in the perturbations and in the equilibrium state (eq. 6), the
differences in $\rho$ are still dominated by the differences in $v$,
and the argument goes through as before. 

We conclude that {\it isentropic, fully catalyzed matter in a 
gravitational field is never unstable to convection} (and usually
stably stratified). Of course, the stability is enhanced if there
is an additional, positive contribution from the entropy gradient
to the Brunt-V\"ais\"al\"a frequency ($(d S/d P)_{eq}<0$), 
as is the case in an isothermal fluid.

\section{Discussion}

An indirect argument leading to the same conclusion could have been
given by {\it reductio ad absurdum}. Assume that the matter is isentropic,
in chemical equilibrium, but {\it unstable} to convection. Then, convective
overturn interchanges fluid elements initially at different pressures,
%(without changing their uniform entropy), 
which at their new positions are out of chemical equilibrium. 
Over a long enough time, reactions 
reestablish chemical equilibrium, and therefore the initial, 
unstable state. This process leads to a {\it perpetuum mobile}, 
inconsistent with energy conservation. This argument, although 
strongly arguing for the previous conclusion, gives no insight into the local
physics that prevents the described process from occuring, but instead 
provides an additional motivation for the analysis given above. 

Let us now consider the application of the result to matter inside
neutron stars. Immediately after the initial collapse, these stars are
hot and opaque to neutrinos (\cite{Sato,Arnett}), whose non-uniform distribution 
violates the condition of full chemical equilibrium. 
Both this effect and a significant entropy gradient act towards destabilizing 
the star (\cite{Epstein}), leading to vigorous convection (\cite{Keil,Pons}).
However, neutron stars older than a few seconds have lost essentially all
their initial entropy and no longer have any neutrinos trapped 
inside (e.g., \cite{Pons}). At this time, they have settled into an 
essentially zero-entropy, full chemical equilibrium state. The result 
derived in the previous section implies that, regardless of the large 
uncertainties in the equation of state for dense matter (\cite{EOS}),
these stars are stably stratified. Marginal stability is in principle 
possible, but unlikely to hold over extended regions. 
Furthermore, the small remaining entropy is distributed so as to give 
a nearly uniform temperature (\cite{Nomoto}), which also has a stabilizing effect.
The implications of stable stratification for different aspects of neutron 
star astrophysics have been mentioned in the Introduction.

\acknowledgments

The writing of this Letter was motivated by a talk given by C. Thompson within 
the program on {\it Spin and Magnetism in Young Neutron Stars} at the 
Institute for Theoretical Physics in Santa Barbara. Parts of it are
based on discussions with P. Goldreich which took place several
years before. Thanks are also due to J. V\'eliz for preparing the figure.
This work was supported by FONDECYT (Chile) 
under grant no. 8970009 ({\it L\'\i neas Complementarias}), 
and by the National Science Foundation (USA)
under grant no. PHY94-07194.

\clearpage

\begin{figure}
\plotone{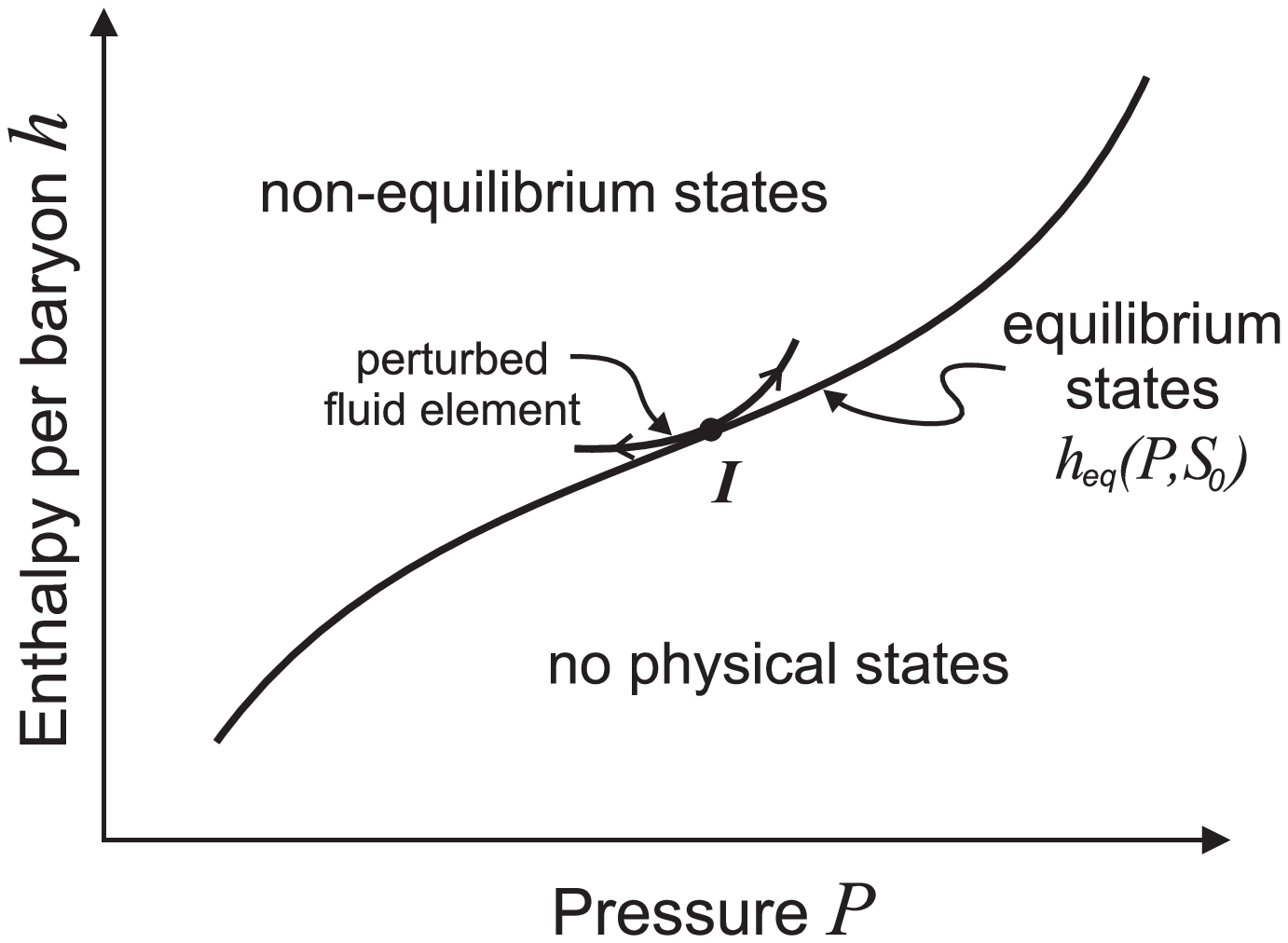}
\caption{Schematic representation of the enthalpy per baryon
as a function of fluid pressure. The long, curved line represents the
equilibrium states (which minimize the enthalpy per baryon at given
pressure and entropy per baryon). Points above this line correspond
to non-equilibrium states, and there are no physical states 
below the line. The shorter line with arrows represents the oscillating 
trajectory of a fluid element slightly perturbed from its equilibrium
position at point $I$. \label{fig1}}
\end{figure}

\clearpage 

\end{document}